\documentclass{article}

\usepackage[nonatbib, final]{tackling_climate_workshop_style}

\usepackage[utf8]{inputenc} 
\usepackage[T1]{fontenc}    
\usepackage{hyperref}       
\usepackage{url}            
\usepackage{booktabs}       
\usepackage{amsfonts}       
\usepackage{nicefrac}       
\usepackage{microtype}      
\usepackage{xcolor}
\usepackage{amsmath}
\usepackage{mathtools}
\usepackage{enumitem}
\usepackage[most]{tcolorbox} 

\definecolor{clean}{RGB}{4,251,4}
\definecolor{hybrid}{RGB}{102,170,155}
\definecolor{dirty}{RGB}{164,84,4}
\title{Towards a Health-Based Power Grid Optimization \\ in the Artificial Intelligence Era}

\author{%
 Claudio Battiloro\thanks{Co-first authors. $\dagger$ Co-last authors. Email: \{cbattiloro@hsph.harvard.edu\}. }, Gianluca Guidi$^*$,\\ \textbf{Falco J. Bargagli Stoffi$^\dagger$, Francesca Dominici$^\dagger$}\\
 Harvard T.H. Chan School of Public Health,
 UCLA Fielding School of Public Health
}

\begin{document}

\maketitle
\begin{abstract}\vspace{-.4cm}
The electric power sector is one of the largest contributors to greenhouse gas emissions in the world. In recent years, there has been an unprecedented increase in electricity demand driven by the so-called Artificial Intelligence (AI) revolution. Although AI has and will continue to have a transformative impact, its environmental and health impacts are often overlooked. The standard approach to power grid optimization aims to minimize CO$_2$ emissions. In this paper, we propose a new holistic paradigm. Our proposed optimization directly targets the minimization of adverse health outcomes under energy efficiency and emission constraints. We show the first example of an optimal fuel mix allocation problem aiming to minimize the average number of adverse health effects resulting from exposure to hazardous air pollutants with constraints on the average and marginal emissions. We argue that this new health-based power grid optimization is essential to promote truly sustainable technological advances that align both with global climate goals and public health priorities.
\end{abstract}


\section{Introduction}
Carbon emissions are a major driver of climate change, mainly through the greenhouse effect. Most global emissions originate from electricity generation through fossil fuel combustion \cite{UN2024}. Electricity production mostly involves burning coal, oil, or natural gas, releasing carbon dioxide (CO$_2$) and other greenhouse gases (GHGs) that trap heat in the atmosphere, thus intensifying global warming. Reducing fossil fuel dependency is essential for a sustainable energy future \cite{UN2024}.

The artificial intelligence (AI) revolution has significantly increased the demand for electricity, particularly from data centers that support AI models \cite{dhar_carbon_2020-1}. This surge has escalated electricity consumption, straining power resources and grid infrastructure usage\cite{MIT2024}. The International Energy Agency (IEA) estimates that data centers used 240-340 terawatt hours (TWh) in 2022, with consumption set to double by 2026 \cite{IEA_2024}.

Power grids, which distribute electricity, now face pressure to optimize operations to meet AI-driven electricity needs while minimizing costs, environmental impacts, and CO$_2$ emissions. Standard approaches focus on multi-objective optimization, either minimizing CO$_2$ emissions (emission-based) or improving energy efficiency (efficiency-based). See Appendix A for a brief literature review. 

However, we advocate for a broader approach, dubbed as \textbf{\textit{health-based power grid optimization}}, that explicitly takes into account health impacts from emissions, including fine particulate matter (PM$_{2.5}$) and hazardous air pollutants (HAPs), which negatively affect human health. As AI increases electricity demand, this holistic paradigm addresses both environmental and health consequences, essential for sustainable grid optimization in the AI era.

\begin{figure}[t!]
  \centering
\includegraphics[keepaspectratio, width=1\textwidth]{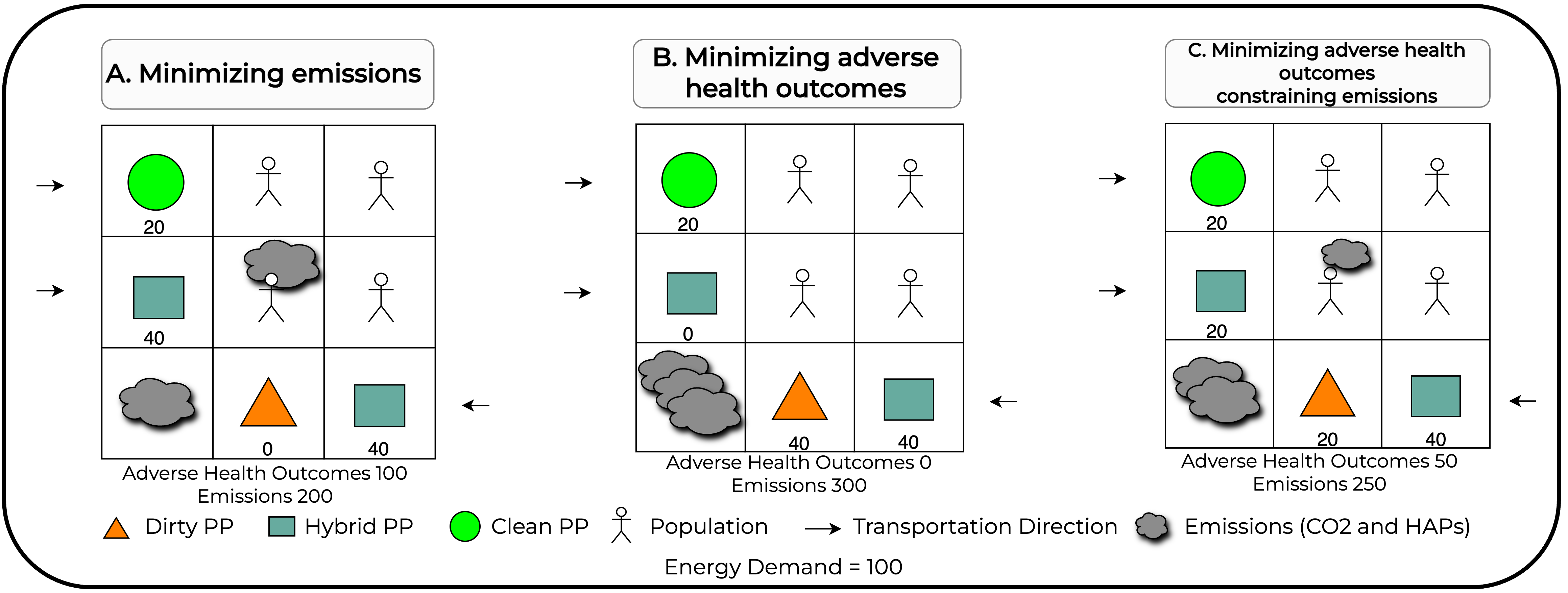}
  \caption{\small A toy example showing that minimizing emissions is not the same as minimizing adverse health outcomes, and the need for health-based power grid optimization. Assume a grid world where each square represents an area with either a power plant (\textcolor{clean}{clean}, \textcolor{hybrid}{hybrid}, or \textcolor{dirty}{dirty}) or a population. \textcolor{clean}{Clean}, \textcolor{hybrid}{hybrid}, and \textcolor{dirty}{dirty} plants can produce, with the available supply, 20, 40, and 40 units of energy, respectively. Assume the energy demand is 100, and the weather-dependent emission transport directions are as depicted by the arrows. In A, energy demand is allocated to power plants by minimizing emissions, therefore \textcolor{clean}{clean} and \textcolor{hybrid}{hybrid} plants are used. However, emissions reach populated areas, causing significant adverse health outcomes. In B, energy demand is allocated to power plants to minimize health impacts, therefore \textcolor{clean}{clean}, \textcolor{hybrid}{hybrid}, and \textcolor{dirty}{dirty} plants are used while avoiding emissions in populated areas. However, the overall emissions are high. In C, our framework, energy is allocated to power plants to minimize adverse health outcomes while constraining emissions, achieving lower emissions than B and lower adverse health outcomes than A.}
  \label{fig:}
\end{figure}
\section{A Pathway to Health \& Climate Impact via Power Grid Optimization}
Our goal is to minimize the downstream health impacts of the increasing electricity demand, under emission constraints. Achieving this goal requires fundamental changes in the current methodologies. In this paper, we propose, for the first time, a paradigm shift from efficiency-based and emission-based to health-based power grid optimization. The key motivation is the following: 
\begin{tcolorbox}[colback=lightgray!5, 
                  colframe=black!40,     
                  arc=10pt,             
                  boxrule=.5mm,          
                  width=\textwidth,     
                  top=1pt,             
                  bottom=1pt,          
                  left=7pt,            
                  right=7pt]           
\textbf{Motivation.} Although adverse health outcomes are a \textbf{monotonic function} of exposure to HAPs, minimizing emissions tout-court \textbf{is not} the same as minimizing adverse health outcomes. This is because a reduction in HAPs emissions is \textbf{not related 1-to-1} to a reduction in CO$_2$ emissions. Furthermore, \textbf{the strength and slope} of the relationships between HAPs and adverse health outcomes vary across regions and \textbf{account} for air pollution transport and energy demand.
\end{tcolorbox}
The consequence of this is that minimizing emissions tout-court does not always result in minimizing adverse health outcomes, and vice versa. In Figure \ref{fig:}, we show a toy example to highlight this fact. Therefore, our proposal pushes for an explicit inclusion of health aspects into power grid optimization: 
\begin{tcolorbox}[colback=lightgray!5, 
                  colframe=black!40,     
                  arc=10pt,             
                  boxrule=.5mm,          
                  width=\textwidth,     
                  top=1pt,             
                  bottom=1pt,          
                  left=7pt,            
                  right=7pt]           
\textbf{Proposal.} Optimization \textbf{should not} be driven by the minimization of energy inefficiency or emissions for the \textbf{implicit mitigation} of adverse health outcomes, but rather by the \textbf{explicit minimization of adverse health outcomes} under energy efficiency and emission constraints.
\end{tcolorbox} Our ultimate goal is soliciting the development of reliable tools that assist policymakers in setting long-term carbon emission caps while also minimizing the health effects caused by HAPs resulting from the power grid. In this way, it is possible to control not only the level of carbon emissions but also (and especially) the level of HAPs and the health burden linked to the power grid. At a high level, the optimization of the health-based power grid will provide balanced frameworks that protect public health through environmental sustainability, reflecting the complex challenges of energy demands in the climate crisis in light of the AI revolution. We can leverage a multitude of technical tools to implement health-based power grid optimization. In the next section, we provide the first system model and problem formulation for health-based power grid optimization.

\noindent\textbf{Description of the Proposed Framework.} We now present the first instance of health-based power grid optimization.
Formally, we assume to have a spatial region $\mathcal{S}$ (e.g., the US) tasseled in $S$ subregions $\{\mathcal{S}_i\}_{i=1}^S$ (e.g., counties). In Figure \ref{fig:}, $S = 9$. Moreover, we assume the allocation to synchronously happen at discrete time steps $t$ (e.g., biweekly).  We assume that power plants in the region $\mathcal{S}$ can be fueled in $N$ different ways (e.g., coal, natural gas, oil, nuclear, solar, wind, hydro, and so on). In Figure \ref{fig:}, we assume $N=3$ (dirty, hybrid, clean). In this setting, the fuel mix of the $i$-th subregion at time $t$, which will be our optimization variable, can be written as a vector $\mathbf{w}_i(t) = [w_{i,1}(t), \dots, w_{i,N}(t)]$ whose entries sum up to one, with the $j$-th entry $0 \leq w_j(t) \leq 1$ being the percentage of energy produced using fuel of the $j$-th type. 


We consider three sources of randomness in each subregion $i$ at time $t$: \textbf{R1} the total energy demand $P_i(t)$, \textbf{R2} the available supplies for each of the $N$ types, and \textbf{R3} the weather conditions $W_i(t)$. We focus on four quantities of interest in each subregion $i$ at time $t$: \textbf{Q1} CO$_2$ emissions $e^{CO_2}_i(t, \mathbf{w}_i(t), P_i(t))$, being a function of the fuel mix and the energy demand, \textbf{Q2} HAPs emissions $e^{HAP}_i(t, \mathbf{w}_i(t), \{W_i(t)\}_{i=1}^S, P_i(t))$, being a function of the fuel mix, the weather conditions (of all the regions, as HAPs can be transported) and the energy demand, \textbf{Q3} the energy availability of each of the $N$ types \{$p_i^n(t, S^n_i(t))\}_{n=1}^N$, being a function of the available supplies, and \textbf{Q4} the number of hospitalizations of HAP-related diseases $h_i(t, e^{HAP}_i(t))$, being a function of the HAPs emissions.

With this system model, we can formulate the first example of a health-based power grid optimization problem. The aim is to find the optimal dynamic fuel mix allocation policy to minimize the average number of hospitalizations due to exposure to HAP with constraints on the average emissions. For the sake of exposition, let us neglect the dependencies among the different variables described above. Then, the dynamic allocation problem can be cast as:
\begin{align}\label{Dyn_Problem}
&\min_{\{\mathbf{w}_i(t)\}_{i=1}^S}\;\; \lim_{t\to \infty} \frac{1}{t}\sum\nolimits_{\tau=0}^{t-1}\mathbb{E} \left \{\sum_{i =1}^Sh_i(\tau)\right\} \nonumber\\
&\text{subject to} \quad \textrm{\textbf{C1}}\; \lim\nolimits_{t\to \infty} \frac{1}{t}\sum_{\tau=0}^{t-1} \mathbb{E}\left\{\sum_{i=1}^S e^{CO_2}(\tau)\right\} \leq C^{CO_2};  \\
& \;\;\;\quad\quad\quad\quad \textrm{\textbf{C2}}\; \lim\nolimits_{t\to \infty} \frac{1}{t}\sum_{\tau=0}^{t-1} \mathbb{E}\left\{\sum_{i=1}^S e^{HAP}(\tau)\right\} \leq C^{HAP}; \quad \textrm{\textbf{C3}}\; 0\leq w_{i,n}(t) \leq 1, \nonumber \\
&  \;\;\;\quad\quad\quad\quad \textrm{\textbf{C4}}\; w_{i,n}(t)P_i(t) \leq p_i^n(t), \quad \textrm{\textbf{C5}}\; \sum_{n=1}^N w_{i,n}(t) = 1, \quad i =1,\dots,S, \quad n =1,\dots,N, \nonumber 
\end{align}
where the expectations are taken with respect to the unknown joint distribution of all the sources of randomness described above. The constraints of \eqref{Dyn_Problem} have the following meaning: \textrm{\textbf{C1}} the average total CO$_2$ emissions do not exceed a threshold/cap value $C^{CO_2}$; \textrm{\textbf{C2}} the average total HAPs emissions do not exceed a threshold/cap value $C^{HAP}$; finally, the instantaneous constraints \textrm{\textbf{C3}}-\textrm{\textbf{C4}}-\textrm{\textbf{C5}} impose that fuel mix is a proper one (\textrm{\textbf{C3}}-\textrm{\textbf{C5}}, the positive fractions must sum up to one), and that the energy demand is met (\textrm{\textbf{C5}}, the allocated energy should be available for all fuel types). 

\noindent\textbf{Discussion and Challenges.} Solving the problem in \eqref{Dyn_Problem} requires striking the best trade-off between minimizing adverse health outcomes, minimizing emissions, and fulfilling the energy demand. The policies in A and B in Figure \ref{fig:} cannot be solutions of \eqref{Dyn_Problem} for reasonable threshold values, because they would neither minimize adverse health outcomes nor meet emission restrictions, respectively. 
Based on the available prior knowledge---e.g., the availability and precision of dispersion models for HAPs emission or of health impact models for linking HAPs and the number of hospitalizations---, and data, the problem in \eqref{Dyn_Problem} can be tackled with hybrid model-based and data-driven approaches, e.g., Lyapunov optimization or model-based reinforcement learning\cite{neely2010lyapunov,battiloro2022FL,moerland2022mbrlsurvey}, or with fully data-driven approaches, e.g. plain deep reinforcement learning \cite{zhang2020rlpower}. In general, the problem in \eqref{Dyn_Problem} can be arbitrarily enriched based on specific needs and available information. Finally, once the optimal fuel mix allocation policy is found, it must be implemented, requiring an inner optimization loop to allocate the energy production of single power plants in each subregion.

\noindent\textbf{Conclusion.} Health-based power grid optimization will help to ensure that the benefits of AI do not come at the expense of health. Our framework will allow us to mitigate the negative impacts on public health while addressing environmental challenges. Our goal is to foster a more balanced and informed conversation about the role of AI in the Climate Crisis, seen as a Health Crisis, highlighting the interconnectedness of these global issues and the need for responsible innovation.

\bibliographystyle{IEEEbib}
\bibliography{ref}

\begin{thebibliography}{10}
\bibitem{MIT2024}
``We’re getting a better idea of {AI}’s true carbon footprint,'' .
\end{thebibliography}


\begin{appendix}
\section{Related Work}
Previous studies have investigated the optimization of power grids, typically focusing on either improving the efficiency of electricity distribution (efficiency-based), or mitigating environmental impacts by minimizing carbon emissions (emission-based). 

\medskip

In terms of efficiency-based methods, a recent survey\cite{motta2024survey} highlights various approaches that aim to improve energy supply efficiency by reducing costs or optimizing energy transmission. These efforts have mainly focused on three aspects of power system planning: (i) \textit{optimal power flow}\cite{momoh1999review}, (ii) \textit{unit commitment}\cite{bingane2018tight}\cite{tejada2019unit}, and (iii) \textit{generation and transmission expansion planning}\cite{hemmati2013comprehensive, meza2007model,unsihuay2011multistage}.

\medskip

Regarding emission-based methods, a significant body of literature\cite{hashim2005optimization, jenkins2018getting,chen2021distribution,kang2020optimizing,boffino2019two,liu2022cutting} concentrated on optimizing the power grid to reduce carbon emissions and, consequently, environmental impacts. However, despite extensive research in these areas, none of the reported studies have considered optimizing the power grid prioritizing the health impacts. This shortcoming represents a significant gap in the literature, as the health consequences of power generation and distribution---such as air pollution and its associated morbidity and mortality---are crucial factors that should be integrated into grid optimization models. Addressing this gap could lead to more comprehensive and socially beneficial approaches to power grid management.
\end{appendix}
\end{document}